\documentclass[10pt]{article}
\usepackage[top=2cm,left=2cm,right=2cm]{geometry}
\usepackage[numbers,square]{natbib}

\usepackage[english]{babel}
\usepackage{epsfig}
\usepackage{hyperref}
\usepackage{amssymb}
\usepackage{amsfonts}
\usepackage{epsf}
\usepackage{rotating}
\usepackage{graphicx}
\usepackage{amsmath}
\usepackage{fancyhdr}
\usepackage{lineno}
\usepackage{subfigure}
\usepackage{babel}
\usepackage{graphics}
\usepackage{geometry}
\usepackage{pstricks}
\usepackage{color}

\newcommand{\xu}{x_{1}}

\newcommand{\beqa}{\begin{eqnarray}}
\newcommand{\eeqa}{\end{eqnarray}}
\newcommand{\bea}{\begin{eqnarray}}
\newcommand{\eea}{\end{eqnarray}}
\newcommand{\beq}{\begin{equation}}
\newcommand{\eeq}{\end{equation}}

\newcommand{\pd}{\partial}

\newcommand{\nn}{\nonumber}

\begin{document}

\begin{center}
\vspace{4.cm}
{\bf \large  Conformal anomaly actions for dilaton interactions \footnote{Presented at \emph{QCD@Work 2014}, 16-19 June 2014, Giovinazzo (Bari - Italy)}}

\vspace{1cm}

{\bf $^a$Luigi Delle Rose, $^a$Carlo Marzo and $^a$Mirko Serino}

\vspace{1cm}

{\it$^a$ Dipartimento di Matematica e Fisica \\
Universit\`a del Salento\\ and \\
INFN Lecce, Via Arnesano 73100 Lecce, Italy \footnote{luigi.dellerose@le.infn.it, carlo.marzo@le.infn.it, mirko.serino@le.infn.it}}

\end{center}
 
\begin{abstract}
We discuss, in conformally invariant field theories such as QCD with massless fermions, a possible link between the perturbative signature of the conformal anomaly,  
in the form of anomaly poles of the 1-particle irreducible effective action, and its description in terms of Wess-Zumino actions with a dilaton. 
The two descriptions are expected to capture the UV and IR behaviour of the conformal anomaly, in terms of fundamental and effective degrees of freedom respectively, 
with the dilaton effective state appearing in a nonlinear realization. 
As in the chiral case, conformal anomalies seem to be related to the appearance of these effective interactions in the 1PI action 
in all the gauge-invariant sectors of the Standard Model. 
We show that, as a consequence of the underlying anomalous symmetry, the infinite hierarchy of recurrence relations 
involving self-interactions of the dilaton is entirely determined only by the first four of them. 
This relation can be generalized to any even space-time dimension. 
\end{abstract}

%

\section{Introduction}
Conformal anomalies play a significant role in theories which are classically scale invariant, such as QCD. 
This type of anomaly manifests as a non-vanishing trace
of the vacuum expectation value (vev) of the energy momentum tensor in a metric ($g_{\mu\nu}$) and gauge field ($A_\mu$) background. If the spontaneous breaking of symmetries is associated with massless Goldstone modes of a theory, it is also natural to associate to the anomalous  breaking of the conformal symmetry a massless state, identified as the dilaton. 
Non-perturbative effects may be responsible for the generation of a mass for this state, as 
do radiative corrections in the form of a usual Coleman-Weinberg potential, by a perturbative resummation.   
The anomaly is expressed by the functional relation 
\bea
g_{\mu\nu}\langle T^{\mu\nu} \rangle_s
&\equiv&
\mathcal{A}[g,A] =
\beta_a\, F + \beta_b\, E_4 + \beta_c\,\Box R - \frac{\kappa}{4}  \, F^{\mu\nu}\, F_{\mu\nu} \nn \\
F 
&\equiv&
C^{\alpha\beta\gamma\delta}C_{\alpha\beta\gamma\delta} =
R^{\alpha\beta\gamma\delta}R_{\alpha\beta\gamma\delta} - 2\, R^{\alpha\beta}R_{\alpha\beta} + \frac{1}{3}R^2  \nn \\
E_4 
&\equiv&
R^{\alpha\beta\gamma\delta}R_{\alpha\beta\gamma\delta} - 4\,R^{\alpha\beta}R_{\alpha\beta} + R^2 \, ,
\label{TraceAnomaly}
\eea
with the $\beta$'s, $\kappa$ constants which depend on the particle content of the theory. 
The first two scalars appearing in the anomaly functional given above correspond to the square of the Wey tensor $(F)$ 
and to the Euler-Poincar\'e density ($E_4$), here specialized to 4 space-time dimensions. 

\begin{figure}[h]
\begin{center}
\includegraphics[scale=0.3]{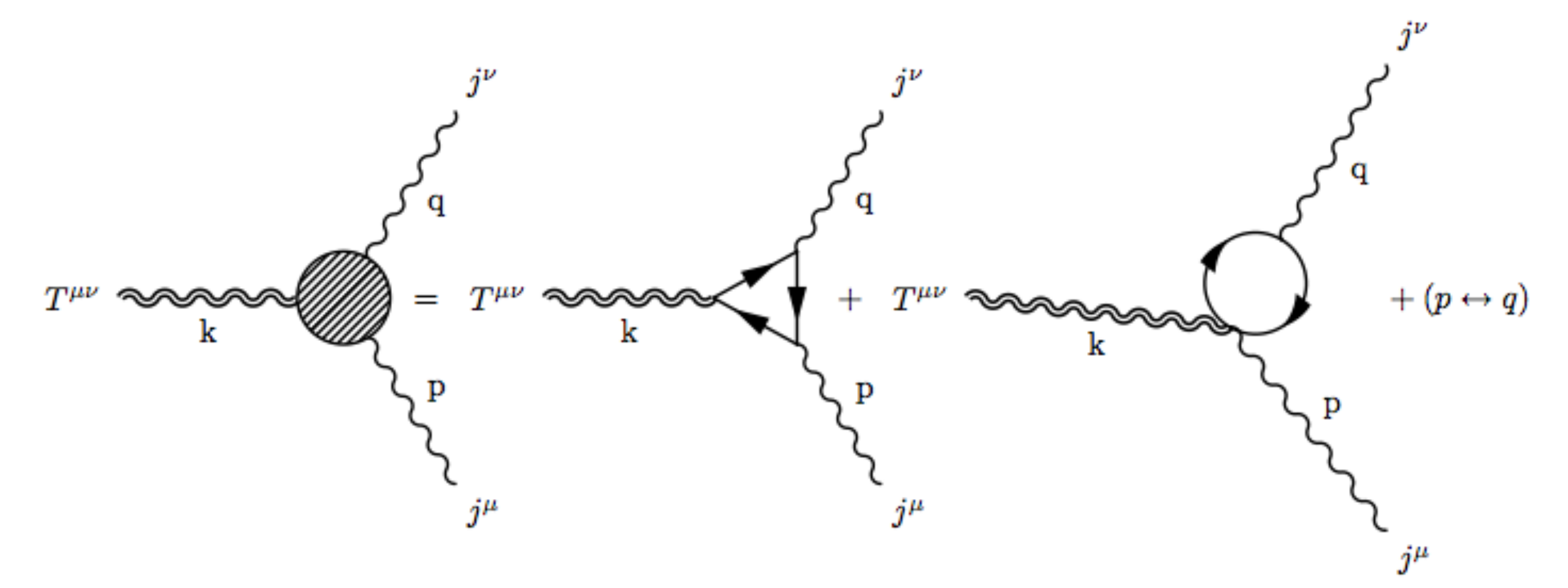}
\caption{Diagrammatic expansions of the $TVV$ vertex}
\end{center}
\label{Fig}
\end{figure}

A perturbative realization of the conformal anomaly is found in the Green function of one energy-momentum tensor and two vector currents, the $TVV$ (see Fig. \ref{Fig}) vertex. We start by discussing this contribution and try to clarify the connection between the perturbative description of the conformal anomaly both at high and low energy, elaborating on the respective effective actions which account for it. We begin from the chiral anomaly, as an example, moving then to 
the conformal anomaly at a second stage. \\
Direct perturbative computations \cite{Dolgov:1971ri, Armillis:2009sm} show that, in momentum space, for massless fermions 
and on-shell photons of momenta $k_1$ and $k_2$, the chiral anomaly diagram reduces to a single tensor structure 

\bea
\Delta^{\lambda\mu\nu}(k_1,k_2) &=& 
\frac{i\,Q^2}{2\, \pi^2}\, \frac{k^\lambda}{k^2}\,\epsilon^{\mu\nu\alpha\beta}\, k_{1\,\alpha}\, k_{2\,\beta},
\label{delta}
\eea
with $k =k_1+k_2$, where we have denoted with $\lambda$ the index associated with the axial vector current and with $\mu,\nu$ those of the two vector currents. 
In this case the form factor is trivially given by a $1/k^2$ term in the axial-vector channel, the anomaly pole  \cite{Dolgov:1971ri}. A discussion of the nonlocal anomaly action identified by this contribution in position space can be found in \cite{Armillis:2009sm, Coriano:2008pg, Armillis:2011hj}. 
Obviously, it is by now a common lore that the anomaly pole appearing in the perturbative computation is matched, in the chiral limit of QCD, to the bound state of a quark-antiquark pair, identified with the pion. In this respect, the perturbative pole of the 
chiral anomaly provides a window on the non perturbative phase of QCD, with the anomaly reproduced, at low energy, by a composite state via a nonlinear (but local) Wess-Zumino action. This second form of the anomaly action carries the same anomaly content of the 1-particle irreducible effective action (1PI) obtained from (\ref{delta}), although the passage from the nonlocal to the local formulation is, obviously, a complicated and unsolved problem, related to confinement and the vacuum structure of QCD. There is no doubt, though, that the appearance of poles in the 1PI action is an important perturbative aspect which requires a close attention. \\
  It is also clear, though, that the anomaly action is not unique. Both the local and the nonlocal versions of this action are legitimate descriptions of the same phenomenon, and we should use, for phenomenological applications,  
the version of the two which is the most appropriate. The nonlocal anomaly action, for instance, identified via the anomaly pole, is describing the conformal anomaly from a perturbative perspective and corresponds to a UV description. Its counterpart, at lower energy, is a local action where the interpolating massless state appears as an asymptotic field.  We just recall that Wess-Zumino actions, introduced as a variational solution of the anomaly condition, carry no reference to perturbation theory and introduce an extra field, which shifts as a Goldstone 
mode under the symmetry. This field describes the same composite state previously identified in the UV as a simple virtual exchange in the Feynman expansion. 
This field can be a scalar or a pseudoscalar. In the case of the conformal anomaly this scalar is the dilaton.\\
As we have emphasized above,  anomaly poles are not only a feature of chiral anomalies, but also of trace anomalies, 
as shown from the computation of the spectral densities of the $TVV$ correlator in pure QED \cite{Giannotti:2008cv, Armillis:2009pq}, 
QCD and electroweak theory \cite{Armillis:2010qk, Coriano:2011zk}.  
In massless QED, for instance, which is a conformally invariant theory affected by the conformal anomaly, this is quite evident from the structure of the $TVV$ correlator in momentum space,
\bea
\Gamma^{\mu\nu\alpha\beta}  (k^2,0,0)
&=&
- \frac{e^2}{48 \pi^2} \bigg\{\frac{1}{k^2}\, \left[
\left(2 \, p^\beta \, q^\alpha - k^2 \, g^{\alpha\beta}\right)\, \left( 2 \, p^\mu \, p^\nu + 2 \, q^\mu \, q^\nu - k^2 \, g^{\mu\nu} \right)\right]
\nn \\
&& \hspace{-35mm}
+\, \frac{1}{3 }\bigg( 12\, \log\left(\frac{k^2}{\mu^2}\right) -35 \bigg)\, 
\bigg[ \big(p^{\mu} q^{\nu} + p^{\nu} q^{\mu}\big)\eta^{\alpha\beta}
+ \frac{k^2}{2} \big(\eta^{\alpha\nu} \eta^{\beta\mu} + \eta^{\alpha\mu} \eta^{\beta\nu}\big) 
\nn \\
&& \hspace{-35mm}
-\, \eta^{\mu\nu} (\frac{k^2}{2} \eta^{\alpha \beta}- q^{\alpha} p^{\beta})
-\big(\eta^{\beta\nu} p^{\mu} + \eta^{\beta\mu} p^{\nu}\big)q^{\alpha} 
- \big (\eta^{\alpha\nu} q^{\mu} + \eta^{\alpha\mu} q^{\nu }\big)p^{\beta} \bigg] \bigg\},
\label{ap}
\eea
which exhibits again the $1/k^2$ behaviour of the anomaly, as in the chiral case. Here $k$ is the momentum of the graviton and $p$ and $q$ those of the two photons. This pole-like behaviour, as shown in  
\cite{Coriano:2012nm}, is naturally inherited by the dilatation current $J_D=x \,T$ of the  conformal theory. Obviously, given this perturbative evidence, which we interpret as an ultraviolet description, it is then natural to thread our way, also in this case, by following the example of the chiral anomaly. If the anomaly pole is the signature of the pseudo-Goldstone boson of the broken conformal symmetry, then we should look for an action which reproduces the same anomaly and is suitable for the description of this virtual exchange as a composite asymptotic field. This action will describe a composite \emph{dilaton} and will be characterized by a new scale, $\Lambda$, expected to be larger than the electroweak scale $v$. This perturbative analysis has been carried out in all the neutral currents sectors of the Standard Model \cite{Armillis:2010qk, Coriano:2011zk}, with identical results. \\
We are now going to highlight the construction of the local  anomaly action which is expected to capture the dynamics of this composite state using a special procedure called {\em Weyl gauging}. As in the chiral case, also in this case it is natural to expect that all the anomalous interactions of the effective dilaton will be derived from the anomaly functional. 
For instance, the part of the local effective action describing the anomaly pole appearing in (\ref{ap}) is trivial to write down 
\bea
\Gamma[A_\mu,\tau] &=& \int d^4x\, \frac{\tau}{\Lambda}\, F^{\mu\nu}F_{\mu\nu} + \dots (\text{mass terms}) \, ,
\label{DilatonGauge}
\eea
with $\Lambda$ the conformal scale, as the field strength $F_{\mu\nu}$ is inert under Weyl scaling. 
In general, however, the derivation of the remaining contributions requires considerable effort, 
also because of the contributions coming from the {\em local anomaly} terms which are prescription dependent and that we prefer to keep. 
As in the chiral case, the development of a step-by step 
procedure which would allow to move from one description to the other remains an open issue.

\section{Weyl-gauging and the Wess-Zumino anomaly action for the dilaton}
To derive the local dilaton anomaly action, we have extended the method of {\em Weyl-gauging}, first proposed for classical theories in \cite{Iorio:1996ad} (see also \cite{Codello:2012sn})  to the renormalized (quantum) action responsible for the conformal anomaly.   
The general strategy for making a classical field theory Weyl invariant involves field-enlarging transformations such as
\bea
 \Phi 
&\to & 
\Phi \, e^{d_{\Phi}\tau/\Lambda}\, , \nn \\
\nabla_\mu 
&\to& 
\nabla_\mu + \left( - d_\Phi\, {\delta^\nu}_\mu + 2\, {\Sigma^\nu}_\mu \right)\, W_\nu, \nn\\
\hat{g}_{\mu\nu}&\to& g_{\mu\nu}e^{-2\tau/\Lambda}\, ,
\eea
with $\tau$ being the dilaton and $W_\mu$ an abelian gauge vector field. Under a Weyl transformation with a parameter $\sigma$ , the change in the metric is given by
\bea
\hat{g}_{\mu\nu} &\to& \hat{g}_{\mu\nu} \equiv g_{\mu\nu}\,  e^{2\,\sigma(x)}\, ,
\eea
while the Weyl field $W_{\mu}$ transforms as an ordinary abelian connection  
\bea
W_{\mu} &\to& W_{\mu} + \pd_\mu \sigma \, .
\label{WeylTransf}
\eea
For a theory with a scale invariant Lagrangean, the compensator field $\tau$ does not appear explicitly. However, if the theory is not scale invariant, for instance due to the presence of dimensional parameters ($\mu$), there is no way of avoiding the introduction of a compensator field. The gauging procedure renders the general dimensionful parameter $\mu$ Weyl invariant via the field enlarging transformation 
\bea
\mu &\to& \mu\, e^{- d_{\mu} \tau/\Lambda} 	\, .
\eea
Under a Weyl transformation, the transformation rule for $\tau$ is that of a Goldstone mode, with a local shift 
\bea
 \tau &\to& \tau - \Lambda\, \sigma(x) \, .
\label{ShiftDilaton}
\eea
At a second stage the compensator is made dynamical with the inclusion of a kinetic term \cite{Coriano:2013nja}. 
In a minimal description, the Weyl gauge connection is traded for a derivative of the dilaton field
\bea
W_\mu &\equiv& \frac{\pd_\mu \tau}{\Lambda} \, ,
\eea
which has significant aesthetic appealing, since the gradient of the dilaton field takes the role of a gauge connection for the implementation of the Weyl symmetry. Below we will provide few more details concerning this construction, which exemplifies the approach to be followed in the computation of the same action for realistic situations, the Standard Model being the most important case. 
In general, there can be two kinds of contributions to the dilaton effective action:
\begin{enumerate}
\item diffeomorphism$\times$ Weyl-invariant terms, carrying no information on the anomalous interactions;
\item diffeomorphism-invariant $\times$Weyl-non-invariant terms, describing anomalous interactions.
\end{enumerate}
The first kind of contributions are easy to identify, as they are simply given by all the terms
obtained by the Weyl-gauging of all the diffeomorphic invariant functionals \cite{Coriano:2013xua},
\bea
\mathcal{J}_n &\sim& \frac{1}{\Lambda^{2(n-2)}}\int d^4 x \sqrt{\hat{g}}\, \hat{R}^n\, .
\eea
It is interesting to investigate the effect of Weyl gauging on the non Weyl-invariant part of the quantum effective action. This is best done by working in dimensional regularization, because in such a scheme the terms in the trace anomaly are in a one to one correspondence
with the 1-loop counterterms \cite{Duff:1977ay}. Denoting with $\Gamma_0[g]$ the classical unrenormalized action and with ${\Gamma}_{\textrm{Ct}}[g] $ its counterterm, the renormalized action $\Gamma[g]$ takes the form 
\bea
\Gamma[g]  &=& 
\Gamma_0[g] + {\Gamma}_{\textrm{Ct}}[g],
\eea
with
\bea
{\Gamma}_{\textrm{Ct}}[g] 
&=& 
- \frac{\mu^{-\epsilon}}{\epsilon}\int d^d x\, \sqrt{g}\, \bigg( \beta_a F + \beta_b E_4\bigg) \, ,  \quad \epsilon = 4 - d.  \eea
The anomaly is generated by the counterterm 
\bea
g_{\mu\nu}\frac{\delta{\Gamma}_0[g]}{\delta g_{\mu\nu}}\bigg|_{d\rightarrow 4} &=& 0  \, ,
\quad
\frac{2}{\sqrt{g}}g_{\mu\nu}\frac{\delta{\Gamma_{\textrm{Ct}}}[g]}{\delta g_{\mu\nu}}\bigg|_{d\rightarrow 4} =
\beta_a\, \bigg( F - \frac{2}{3}\Box R \bigg)+ \beta_b\, E_4 \, .
\eea
In the case of the square of the Weyl tensor $F$, for instance, starting from the expansion of the Weyl-gauged counterterms (with everything computed in $d$ dimensions) one derives the functional relation
\bea
- \frac{1}{\epsilon}\,\int d^dx \, \sqrt{\hat{g}} \, \hat{F} 
&=& 
- \frac{1}{\epsilon}\, \int d^dx \sum_{i,j=0}^{\infty}\frac{1}{i!j!}\, \epsilon^i\,  \frac{1}{\Lambda^j} \,
\frac{\pd^{i+j}\left[\sqrt{\hat g}\, \hat F\, \right]}{\pd \epsilon^i\,\pd (1/\Lambda)^{j}},
\label{PreGaugingF}
\eea
which plays a key role in the derivation of the local version of the anomaly action.
Obviously, only the $O(\epsilon)$ contributions in the power-series expansion above are significant, due to the $1/\epsilon$ pole. A similar relation holds for the $E_4$ counterterm.
It is clear that by a Weyl gauging of the counterterm action we are generating an action which is Weyl invariant
\bea
\hat\Gamma_{\textrm{ren}}[g,\tau] 
\label{one}
&=&
\Gamma_{\textrm{ren}}[g,\tau] - \Gamma_{WZ}[g,\tau]  \, , \\
\delta_W \hat\Gamma_{\textrm{ren}}[g,\tau]  &=& 0 \, , 
\eea
with $\delta_W$ the Weyl variation, showing that the new terms generated by this procedure correspond to the dilaton Wess-Zumino action (up to a sign).
After rearranging (\ref{PreGaugingF}) and the analogue equation for $\int d^dx\, \sqrt{g}\, \hat{E_4}$ through many integrations by parts, we get the final expression of the dilaton action
\bea
&&
\Gamma_{WZ}[g,\tau] =
\Gamma_{\textrm{ren}}[g,\tau] - \hat\Gamma_{\textrm{ren}}[g,\tau] =
\int d^4x\, \sqrt{g}\, \bigg\{ \nn \\
&&
\beta_a\, \bigg[ \frac{\tau}{\Lambda}\, \bigg( F - \frac{2}{3} \Box R \bigg) + 
\frac{2}{\Lambda^2}\, \bigg( \frac{R}{3}\, 	\left(\pd\tau\right)^2 + \left( \Box \tau \right)^2 \bigg)
-\, \frac{4}{\Lambda^3}\, \left(\pd\tau\right)^2 \,\Box \tau + \frac{2}{\Lambda^4}\, \left(\pd\tau\right)^4 \bigg] \nn \\
&& \hspace{-3mm}
+\, \beta_{b}\, \bigg[ \frac{\tau}{\Lambda}\,E_4  - 
\frac{4}{\Lambda^2}\, \bigg( R^{\alpha\beta} - \frac{R}{2}\,g^{\alpha\beta} \bigg)\, \pd_\alpha\tau\, \pd_\beta\tau - 
\frac{4}{\Lambda^3} \, \left(\pd\tau\right)^2 \,\Box \tau +  \frac{2}{\Lambda^4}\, \left( \pd\tau\right)^4 \bigg] \bigg\} \, .
\eea
At the price of introducing one extra degree of freedom, the dilaton, we have naturally obtained a Weyl-invariant quantum effective action, represented by $\Gamma_{\textrm{ren}}[g,\tau]$. Obviously, at this stage, we view this result just as a formal trick which allows to extract
$\Gamma_{WZ}[g,\tau] $ in an interesting way and not as an attempt to erase the conformal anomaly at quantum level. In particular, 
in the flat space limit (with $g\to \delta$) we have the result
\bea
\Gamma_{WZ}[\delta,\tau] 
&=&
\int d^4x\,  \bigg[
\frac{2\,\beta_a}{\Lambda^2}\, \left( \Box \tau \right)^2
+ \left(\beta_a + \beta_b\right)\, 
\bigg( - \frac{4}{\Lambda^3}\, (\pd\tau)^2 \, \Box \tau + \frac{2}{\Lambda^4}\, \left(\pd\tau\right)^4\, \bigg) \bigg] \, ,
\label{WZFlat}
\eea
which can be very powerful from the computational point of view.

In fact, from (\ref{WZFlat}) it is easy to extract the dilaton interactions predicted by the anomaly $\mathcal I_{n}(\xu,\dots,x_n)$, defined as
\bea
\mathcal I_{n}(\xu,\dots,x_n) &=& 
\frac{\delta^n \left(\hat\Gamma_{\textrm{ren}}[\delta,\tau]-\Gamma_{\textrm{ren}}[\delta,\tau]\right)}
{\delta\tau(\xu)\dots\delta\tau(x_n)} = -  \frac{\delta^n \Gamma_{WZ}[\delta,\tau]}{\delta\tau(\xu)\dots\delta\tau(x_n)} \, .
\label{Vertices}
\eea
A remarkable feature of (\ref{Vertices}) is that
\bea
\mathcal I_{n}(\xu,\dots,x_n) &=& 0 \, , \quad n \geq 5 \, ,
\eea
a constraint which can be generalized to any conformal theory in any even dimension $d$, showing that there are 
no dilaton interactions for $n>d$. \\
Thanks to this relation, we can draw some significant constraints on the traced correlators of the energy-momentum tensor for classically
conformally invariant field theories.
This program has been completed in $2$ and $4$ dimensions in \cite{Coriano:2013xua} and for $6$ dimensions in \cite{Coriano:2013nja}.

\section{Conclusions} 
The anomalous breaking of a conformal symmetry manifests in the appearance of a dilaton in the effective action. 
We have elaborated on the fact that the signature of this breaking, according to the ultraviolet description provided by the Feynman expansion, consists in the appearance of anomaly poles (found in 
all the gauge invariant sectors of QCD as of the electroweak theory) in those correlators which manifest the anomaly.  
According to the perturbative picture, the corresponding action takes a nonlocal form. On the 
other hand, the local action which describes the dynamics of this effective degree of freedom should take a typical Wess-Zumino form. In our approach we have derived this action using a Weyl 
symmetric construction. This allows to obtain, as an important by-product, some nontrivial constraints concerning the structure of the dilaton self-interactions. We expect that the (local) nonlinear 
realization of this action will find significant phenomenological applications both in QCD and in the electroweak theory.

 \end{document}